\documentclass[12pt,preprint]{aastex}

\shorttitle{Magnetic Evolution and Temperature Variation}
\shortauthors{Zhang et al.}

\begin{document}

\title{Magnetic Evolution and Temperature Variation in a
Coronal Hole}

\author{Jun Zhang\altaffilmark{1}, Guiping Zhou\altaffilmark{1},
Jingxiu Wang\altaffilmark{1} and Haimin Wang \altaffilmark{2}}


\altaffiltext{1}{National Astronomical Observatories, Chinese
Academy of Sciences, Beijing 100012, China; wjx@ourstar.bao.ac.cn,
zhougp@ourstar.bao.ac.cn, zjun@ourstar.bao.ac.cn}
\altaffiltext{2}{Big Bear Solar Observatory, New Jersey Institute
of Technology, Big Bear City, CA 92314; haimin@flare.njit.edu}

\begin{abstract}

We have explored the magnetic flux evolution and temperature
variation in a coronal-hole region, using Big Bear Solar
Observatory (BBSO) deep magnetograms and {\it SOHO}/EIT images
observed from 2005 October 10 to 14. For comparison, we also investigated
a neighboring quiet region of the Sun. The coronal hole evolved
from its mature stage to its disappearance during the observing
period. We have obtained the following results: (1) When the
coronal hole was well developed on October 10, about 60 \% of the
magnetic flux was positive. The EUV brightness was 420 counts
pixel$^{-1}$, and the coronal temperature, estimated from the line
ratio of the EIT 195 {\AA} and 171 {\AA} images, was 1.07 MK.
(2) On October 14, when the coronal hole had almost disappeared, 51 \%
of the magnetic flux was positive, the EUV radiance was 530
counts pixel$^{-1}$, and the temperature was
1.10 MK. (3) In the neighboring
quiet region, the fraction of positive flux varied between 0.49 and
0.47. The EUV brightness displayed an
irregular variation, with a mean value of 870 counts
pixel$^{-1}$. The temperature was almost constant at
1.11 MK during the five-day observation. Our
results demonstrate that in a coronal hole less imbalance of
the magnetic flux in
opposite polarities leads to stronger EUV brightness and
higher coronal temperatures.

\end{abstract}


\section{INTRODUCTION}

Coronal holes are cool, low-density regions observed both at low
latitudes and at the polar regions of the Sun \citep{chi99}.
They were first observed in white light by
\citet{wal51}, on X-ray plates by \citet{und67}, and on EUV
line spectroheliograms by \citet{ree70}.
Their predominantly unipolar magnetic fields are
open to the interplanetary space \citep{boh77}, giving rise to
high-speed solar-wind streams that can lead to geomagnetic storms
\citep{kri73}. There are three broad categories of coronal holes:
polar, non-polar and transient ones \citep{har02}. It has been
suggested that magnetic reconnection must occur continuously at
the boundary of coronal holes in order to maintain the coronal-hole
integrity \citep{kah02}. The fast solar wind starts flowing
out in magnetic funnels at heights between 5000~km and 20\,000 km
above the photosphere \citep{tu05}.

The electron temperature is clearly an important
parameter in the corona. A detailed assessment of observations
in coronal holes and the deduced temperatures was
published by \citet{hab93}. Electron temperatures
in the corona can be measured with the help of a
magnesium line ratio of a temperature-sensitive pair
\citep[cf.][]{wil06}, with the assumption that the density
and temperature of the gas from which spectral lines are emitted
are constant along the line of sight \citep{hab93}.
It should be mentioned that temperatures in the inner corona
cannot be accurately derived as there are many sources of uncertainty
such as instrument calibration, line-of-sight effects, departure from
ionization balance, and inaccuracies of the atomic data.

In recent years, the coronal temperatures have been intensively
studied since space observations are
ascertained from the {\it YOHKOH} and {\it SOHO} missions
\citep{har92, har94, mos97}. Using the two {\it SOHO} spectrometers,
CDS and SUMER, electron temperatures were measured as a function of
height above the limb in a polar coronal hole \citep{dav98, wil98}.
\citet{dos00} concluded that the emission line ratio increase
in a polar coronal hole was primarily due to an increase of the
electron temperature with height. \citet{mar00} found that the
hydrogen temperature increased only slightly from
1${\times}$10$^{5}$ K to 2${\times}$10$^{5}$ K in the height range
from 12\,000 km to 18\,000 km, and \citet{stu00} presented that with
increasing formation temperature, spectral lines displayed on
average an increasingly stronger blueshift in coronal holes
relative to the quiet Sun at equal heliospheric angle.
Furthermore, \citet{xia04} reported that the bases of coronal
holes seen in chromospheric spectral lines with relatively low
formation temperatures displayed similar properties as normal
quiet-Sun regions. More recently, \citet{wil06} reported that,
in a polar coronal hole region, the electron
temperatures in plumes are 7.5${\times}$10$^{5}$ K and
1.13${\times}$10$^{6}$ K  in inter-plume regions, in the height
of 45 Mm above the limb.

In this Letter, we study the magnetic evolution, the EUV brightness
changes, and the coronal temperature variations in a coronal hole
and an adjacent quiet region from
2005 October 10 to 14. Initially, the coronal hole was well
developed. At the end, the coronal hole had almost disappeared.
Combining deep magnetograms (with a noise level of 2 G) from Big
Bear Solar Observatory (BBSO) with {\it SOHO}/EIT observations, we
unravel the nature of the different magnetic properties and
temperature variations in this coronal hole and the quiet region.

\section{OBSERVATIONS}

From 2005 October 10 to 14, the observational target of BBSO was a
coronal hole and a neighboring quiet region. The target was very
close to the equator, centered at S3$^{\circ}$E28$^{\circ}$
on the 10th, and at S5$^{\circ}$W25$^{\circ}$ on the 14th. The
magnetogram was obtained using the digital vector magnetograph
(DVMG) system mounted on the 25 cm refractor. The DVMG system uses
liquid crystal, a Zeiss filter, and a 12 bit digital camera so
that one can accurately measure small intranetwork magnetic
elements on the order of 2 G. The temporal resolution is 90 s, and
the field of view is 300$''$${\times}$300$''$ (0$''$.6
pixel$^{-1}$). The EUV observations were obtained by the Extreme
Ultraviolet Imaging Telescope (EIT) on board of {\it SOHO}. The
instrument generally observes full-disk EUV images in the coronal
171 {\AA} (Fe IX/X, $\approx$ 1 MK), 195 {\AA} (Fe XII, $\approx$ 1.5
MK), or 284 {\AA} (Fe XIV, $\approx$ 2 MK) passbands. A
detailed description of the instrument is provided by \citet{del95}.

Figure 1 shows a full-disk MDI \citep{sch95} magnetogram
({\it top left frame}),
an EIT 195 {\AA} image ({\it middle left frame}) and a temperature
map ({\it bottom left frame}) derived from ratio of the 195
{\AA} and 171 {\AA} images. High values of this ratio generally
outline magnetically closed field regions. The coronal holes are
clearly defined as dark, i.e., cool regions in
the ratio image both on and off the disk \citep{mos97}. Three
windows on the full-disk images outline the field of view of BBSO
magnetograms. Half of the area is a coronal hole, and
the other half a quiet region. Dashed-lines in the three frames of
the right column separate the coronal hole from the quiet region.

\section{MAGNETIC FIELD EVOLUTION AND TEMPERATURE VARIATION}

Figure 2 shows BBSO magnetograms ({\it left column}) in the region
marked by the small windows in Figure 1. An ephemeral region (ER1)
appeared near 17:49 UTC,
October 11. As ER1 is growing, another smaller ER2 appeared between
the two elements of ER1 (see the magnetogram at 19:29
UTC). The negative element of the smaller ER2 merged into the larger
negative element of ER1, and the positive element of ER2,
canceled with the opposite polarity element of the
ER1. Three hours later, ER2 disappeared. The
interaction of the two ERs is associated with the increased EUV
emission in the right column of Fig. 2. Figure 3 presents the evolution of
magnetic flux density versus time in the upper panel,
and in the middle panel the evolution of EUV emission.
By tracking ER1, we find that it went through four
phases (`P1', `P2', `P3', and `P4') in its evolution. Firstly,
ER1 continuously emerged for 15 h, meanwhile its positive flux
canceled with the pre-existing negative flux. At this
stage, the flux emergence was dominant. In the second phase, the flux
was almost stable, as seen from MDI magnetograms. The third
phase began at 04:24 UTC on October 12. A new ER appeared in
the area, and interacted with ER1. Eight hours later, the new ER could
not be tracked any longer. At this time, the mean flux density in
this area reached the highest value. Finally, low-level emergence and
cancellation intermittently persisted for about one day, before
ER1 faded away. There was quite a
close relationship between the evolution of magnetic
flux density and the variation of the EUV brightness. The coronal
temperature, shown in the bottom panel of Figure 3, deduced from
the ratio of the Fe XII and Fe IX/X channels,
also increased in the first three
phases. However in the last phase, the temperature variation did
not follow that of magnetic flux density.

It is well known that coronal holes lie within a predominantly
unipolar magnetic region, but the solar magnetic field is
never strictly unipolar. \citet{wan92} found
that the minority polarity flux occupied 15 \% to 30 \% of the total
magnetic flux in a coronal hole. However,
we have little knowledge about the
magnetic flux evolution in coronal holes. Here we study the
evolution of the magnetic flux and the variation
of the flux imbalance in
a coronal hole during its decaying stage, and compare them with those
in a quiet region. The top panel of Figure 4 displays the
evolution of the total unsigned flux,
measured from BBSO magnetic field data, in the coronal hole and
the quiet region. We notice that during the five-day observations,
the total fluxes in both regions were
almost stable within the standard uncertainty margin.
This result is confirmed
by seeing-free MDI magnetic field data. The fraction of the
positive flux to the total flux in the two regions is presented in
the next panel. About 58 \% of the
magnetic fields was positive in the coronal hole
on October 10, the day when it was well developed.
During the period in which the
coronal hole gradually disappeared, the fraction of the positive
flux decreased accordingly. The fraction decreased to
51 \% at the end, indicating that the
fluxes of the positive and negative polarities were nearly
balanced. In contrast to the coronal-hole region, we found that
the ratio between positive and negative fluxes in the quiet region
did not change significantly. During the five-day observing
period, the fraction of positive flux varied between 49 \% and 47 \%,
indicating that the fluxes of the two
polarities are approximated balanced all the time.

The EUV brightness and temperature variations of the two
regions are presented in the lower panels of
Figure 4. In the quiet region, the EUV radiation
fluctuates without a trend. The mean value is 870 counts
pixel$^{-1}$. The coronal temperature is almost stable at a
level of 1.11 MK during the five days. In the
coronal hole, although the brightness also fluctuates, there is a
clear ascending trend.  From October 11 to 14, the mean EUV
brightness increased from 420 to 530 counts pixel$^{-1}$, i.e.
a relative increase of about 26 \%. The derived temperature
also increased. On October 10
and 11, the temperature was about 1.07 MK, and on
October 14 it reached 1.10 MK.

\section{CONCLUSIONS AND DISCUSSIONS}

In this Letter, we probe the relationship between the magnetic
field evolution, the EUV brightness changes and the coronal
temperature variations in a coronal hole, and compare it with that of
a neighboring quiet region. In the coronal hole, 58 \% of the
magnetic field was positive while the hole was well
developed. When it almost disappeared, the fraction
of the positive flux decreased to 51 \%. This means that one of the
signatures of decay of a coronal hole is the disappearance of the
flux imbalance. Although the EUV emission fluctuated when the
coronal hole was decaying, it clearly had an increasing trend.
From October 11 to 14, the EUV increased from
420 to 530 counts pixel$^{-1}$, a relative increase of
about 26 \%. The coronal temperature, deduced from the Fe IX/X and
Fe XII line ratio, also increased.
When the hole was well developed, the temperature was about
1.07 MK, and when the coronal hole almost
disappeared, the temperature reached 1.10 MK. In
the quiet region, the magnetic fluxes in both polarities were
always approximately balanced. The EUV radiation fluctuates
slightly with a mean of 870 counts pixel$^{-1}$, and the
temperature was stable at a level of 1.11 MK. By
using a similar line-ratio method, \citet{mos97} presented a
full-disk temperature map. From the map, we deduced a coronal
temperature range was from 1.00 MK to
1.10 MK in coronal holes, and from 1.10 MK to 1.20 MK in quiet
regions. Our coronal temperatures are basically
consistent with those of \citet{mos97}.

By checking the flux transport across the coronal-hole boundary
for this region with MDI data, we do not see a significant
migration of positive or negative fluxes.
The only explanation of the magnetic field evolution then is
that positive network flux canceled with ``hidden" negative
intranetwork (IN) flux, i.e., the magnetic flux which was
too weak to be detected. \citet{zha06} found that
the net IN flux is opposite to that of network flux.
The ephemeral regions, in turn, refurbished the
missing flux. If we assume that at the mature stage of a
coronal hole, there are ten units of total flux, about six unit
are positive and four negative. When it evolved to the decayed stage, two
positive units disappeared
due to cancellation with invisible IN flux. Meanwhile ephemeral regions
provide one additional unit of flux for each of the two polarities.
Consequently, five units each for
the positive and negative fluxes would be detected.

\acknowledgments

We would like to thank the referee, Dr. Klaus Wilhelm,
for valuable comments and useful suggestions.
The authors are indebted to the {\it SOHO}/EIT, MDI and BBSO teams
for providing the data. {\it SOHO} is a project of
international cooperation between ESA and NASA. This work is
supported by the National Natural Science Foundations of China
(G40674081, 10573025 and 10233050), the National Basic Research
Program of China under grant G2006CB806303, and two US NASA grants
(NNG0-6GI19G and NNG0-4GG21G).

\clearpage

\begin{figure}
\epsscale{0.70}
\caption{A full-disk magnetogram from {\it SOHO}/MDI
({\it top left}), an EIT 195 {\AA} image from {\it SOHO}/EIT ({\it
middle left}), and a temperature map ({\it bottom left}) of the
ratio from the EIT 195~{\AA} and 171 {\AA} channels. Three windows
in the three full-disk images outline the field of view of BBSO
magnetograms. The small windows in the right column denote a
sub-area where ephemeral regions appear (see Fig. 2) \label{fig1}}
\end{figure}

\clearpage

\begin{figure}
\caption{Time sequences of BBSO magnetograms ({\it
left}) and EIT 195 {\AA} images ({\it right}) on 2005 October 11.
The arrows show two pairs of ephemeral regions, and the ellipses
encircle their areas. The
field of view is about 20$''$${\times}$20$''$. \label{fig2}}
\end{figure}

\clearpage

\begin{figure}
\epsscale{0.80}
\caption{Magnetic flux density evolution ({\it top})
measured from MDI magnetic field data versus time in the
field of view of Fig. 2. The middle and bottom panels display the
corresponding EUV brightness change and temperature variation, respectively.
Three vertical lines separate the four phases (`P1', `P2', `P3',
and `P4') of ER1. \label{fig3}}
\end{figure}

\clearpage

\begin{figure}
\epsscale{0.70}
\caption{The evolution of the total magnetic flux ({\it
top}) and of the positive magnetic flux fraction ({\it upper
middle}) measured from BBSO magnetograms versus time in the
coronal hole and the quiet region shown in the field of view of
the right column of Fig. 1. The lower panels
display the variations of the EUV emission and temperature changes.
\label{fig4}}
\end{figure}

\end{document}